\documentclass[article,onecolumn,showpacs,showkeys,preprintnumbers,elsart]{revtex4}
\usepackage{makeidx}
\usepackage{amssymb}
\usepackage{amsmath}
\usepackage{graphicx}
\usepackage{dcolumn}
\usepackage{bm}
\usepackage[center]{subfigure}
\usepackage{color}
\usepackage[bf, small]{titlesec}
\usepackage[symbol]{footmisc}

\renewcommand\thesection{\arabic{section}}
\renewcommand\thesubsection{\arabic{section}.\arabic{subsection}}
\titleformat{\section}{\bf\normalsize}{\thesection.\,}{0.5pt}{}
\titlespacing{\section}{0cm}{12pt}{12pt}
\titleformat{\subsection}{\bf\small}{\thesubsection.\enspace}{0.5pt}{}
\titlespacing{\subsection}{0cm}{12pt}{12pt}
\providecommand{\keywords}[1]{\small\textbf{\textit{Keywords---}} #1}

\begin{document}

\title{Dynamical control of solitons in a parity-time-symmetric coupler by
periodic management }
\author{Zhiwei Fan$^{a,}$\thanks{%
Corresponding author.}}
\email{zhiweifan1994@gmail.com}
\author{Boris A. Malomed$^{a,b}$}
\affiliation{$^{a}$Department of Physical Electronics, School of Electrical Engineering,
Faculty of Engineering, Tel Aviv University, Tel Aviv 69978, Israel\\
$^{b}$Center for Light-Matter Interaction, Tel Aviv University, Tel Aviv
69978, Israel}

\begin{abstract}
We consider a dual-core nonlinear waveguide with the parity-time ($\mathcal{%
PT}$) symmetry, realized in the form of equal gain and loss terms carried by
the coupled cores. To expand a previously found stability region for
solitons in this system, and explore possibilities for the development of
dynamical control of the solitons, we introduce \textquotedblleft
management" in the form of periodic sinusoidal variation of the loss-gain
(LG) coefficients, along with synchronous variation of the inter-core
coupling (ICC) constant. This system, which can be realized in optics (in
the temporal and spatial domains alike), features strong robustness when
amplitudes of the variation of the LG and ICC coefficients keep a ratio
equal to that of their constant counterparts, allowing one to find exact
solutions for $\mathcal{PT}$-symmetric solitons. A stability region for the
solitons is identified in terms of the management amplitude and period, as
well as the soliton's amplitude. In the long-period regime, the solitons
evolve adiabatically, making it possible to predict their stability
boundaries in an analytical form. The system keeping the Galilean
invariance, collisions between moving solitons are considered too. Slowly
moving solitons undergo multiple collisions, but eventually separate.
\end{abstract}

\maketitle

\section{Introduction}

One of the most fundamental tenets in physics is the charge-parity-time ($%
\mathcal{CPT}$) symmetry, which holds for all Lorentz-invariant systems
obeying the causality principle \cite{field1,field2}. It implies invariance
of the system with respect to the combined parity transformation, $\mathcal{P%
}$, which reverses the coordinate axes; charge conjugation, $\mathcal{C}$,
which swaps particles and antiparticles; and time reversal, $\mathcal{T}$.
Its reduced forms, such as $\mathcal{PT}$ and $\mathcal{CP}$ symmetries, may
be violated in specific situations, but they also play a profoundly
important role in many physical theories. The usual proof of the presence of
the latter symmetries is performed to Hermitian Hamiltonians, whose
eigenvalues are always real.

However, the invariance of the system with respect to the $\mathcal{PT}$ and
$\mathcal{CP}$ transformations does not imply that the underlying
Hamiltonian must necessarily be Hermitian. Indeed, it was known from some
early examples \cite{1}-\cite{5}, and was then discovered, in the systematic
form, by Bender and Boettcher \cite{bender1,bender2} (see also review \cite%
{bender3} and book \cite{ptqm}) that, in the most general case, Hamiltonians
which commute with the $\mathcal{PT}$ operator may include a dissipative
(anti-Hermitian) term. Such $\mathcal{PT}$-symmetric non-Hermitian
Hamiltonians often include a complex potential, $U(\mathbf{r})$, whose real
and imaginary parts must be, respectively, even and odd functions of spatial
coordinates ($\mathbf{r}$), i.e.,
\begin{equation}
U(-\mathbf{r})=U^{\ast }(\mathbf{r}),  \label{star}
\end{equation}%
where $\ast $ stands for the complex conjugate. Actually, $\mathcal{PT}$%
-symmetric Hamiltonians may admit transformation into Hermitian ones \cite%
{Mostafazadeh,Barash}. A well-established fact is that the spectrum of
Hamiltonians with complex potentials subject to constraint (\ref{star}) is
real below a critical strength of the imaginary part of the potential, at
which the $\mathcal{PT}$ symmetry gets broken, making the system unstable
\cite{breaking} (exceptions in the form of models with \textit{unbreakable} $%
\mathcal{PT}$ symmetry are known too \cite{unbreakable}).

Thus far, the $\mathcal{PT}$ symmetry was not directly realized in quantum
systems with complex potentials. On the other hand, a possibility to realize
it was predicted for classical optical media with symmetrically inserted
gain and loss \cite{theo1}-\cite{review}. This possibility is based on the
commonly known similarity between the quantum-mechanical Schr\"{o}dinger
equation and the propagation equation for optical waveguides, written in the
paraxial approximation. Following these ideas, the $\mathcal{PT}$ symmetry
was experimentally implemented in various optical and photonic systems \cite%
{exp1}-\cite{exci3}. Emulation of the $\mathcal{PT}$ symmetry was also
predicted in atomic Bose-Einstein condensates (BECs), assuming that the gain
may be provided by elements working as matter-wave lasers \cite{Cartarius}.

As concerns the emulation of fundamental properties of quantum systems in
terms of classical optics and in semi-classical BEC, (quasi-) $\mathcal{CP}$
symmetries may be implemented too, in continuous \cite{Flach,CP1,CP2} and
discrete \cite{Hadi} media alike.

The $\mathcal{PT}$ symmetry in an optical waveguide (as well as its $%
\mathcal{CP}$ \ counterpart) may naturally combine with the material Kerr
nonlinearity, giving rise to propagation models based on cubic nonlinear Schr%
\"{o}dinger equations (NLSEs) with the complex potentials subject to
condition (\ref{star}). These models may generate $\mathcal{PT}$-symmetric
solitons, which were addressed in many theoretical works \cite{soliton},
\cite{Konotop}-\cite{Barash-discr}, \cite{unbreakable} (see also reviews
\cite{review1,review2}), and experimentally demonstrated too \cite{exp7}.
Although the presence of the gain and loss makes $\mathcal{PT}$-symmetric
media dissipative, solitons exist in them in continuous families, similar to
the commonly known situation in conservative models \cite{families}, while
usual dissipative solitons exist as isolated solutions (\textit{attractors},
if they are stable) \cite{diss2}.

One of basic settings for the realization of the $\mathcal{PT}$ structure is
provided by dual-core waveguides (couplers), with the gain and loss
separately placed in parallel cores, which are coupled by tunnelling of the
field (light, in optics, or matter waves, in BEC). Stable solitons in
conservative couplers with the Kerr nonlinearity were predicted decades ago.
These solitons may be symmetric or asymmetric with respect to the identical
cores, the symmetry-breaking bifurcation happening at a critical value of
the soliton's total energy/norm (in terms of optics/BEC) \cite%
{Wabnitz,Pare,Maimistov,Wabnitz2}, see also a review in Ref. \cite{Peng}.

A remarkable property of the model of the coupler which includes the cubic
nonlinearity in each core, and the above-mentioned $\mathcal{PT}$-symmetric
terms, in the form of the linear gain and loss in the two cores, is that $%
\mathcal{PT}$-symmetric and antisymmetric solitons not only can be found in
an analytical form, but also their stability region can be identified in an
exact form \cite{Driben,Sukho} (this region is finite for the symmetric
solitons, while antisymmetric ones are completely unstable, although their
instability may be weak). Unlike the conservative counterpart of the system,
asymmetric solitons cannot exist in the presence of the gain and loss,
because asymmetry between components of the soliton in the amplified and
damped cores does not admit establishment of the balance between the gain
and loss.

Expansion of the stability region for $\mathcal{PT}$-symmetric solitons and,
more generally, developing methods for dynamical control of the solitons is
a relevant problem. One potential possibility is suggested by the use of the
\textquotedblleft management" technique, i.e., periodic modulations of the
loss-gain (LG) and inter-core-coupling (ICC) coefficients. In terms of the
conservative model of the nonlinear coupler, the management scheme, which
corresponds to $\gamma _{0}=\gamma _{1}=0$ and $\delta >0$ in Eq. (\ref{uv}%
), see below, was introduced in Ref. \cite{Skinner}, where effects of the
management on symmetric and asymmetric solitons and the transition between
them were studied. Similar management schemes are well known to stabilize
otherwise unstable or fragile solitons in other settings, such as the\textit{%
\ dispersion management} applied to solitons in single-core waveguides (the
local dispersion coefficient with a periodically flipping sign \cite%
{book,Turitsyn}), or the stabilization of two-dimensional solitons (which
are otherwise unstable against the critical collapse \cite{Gadi}) by means
of periodic nonlinearity management \cite{Towers,Kraenkel,Ueda}. In Ref.
\cite{DR} a particular realization of the above-mentioned management format,
implemented as periodic sign change of the LG and ICC coefficients, was
applied to the stabilization of symmetric solitons in the $\mathcal{PT}$-%
\emph{supersymmetric} coupler with the cubic intra-core nonlinearity and
equal LG and ICC coefficients (the supersymmetry implies setting $\gamma
_{0}=1$ in terms of Eq. (\ref{uv}) with $\gamma _{1}=\delta =0$, see below).
In the supersymmetric coupler with constant parameters, all solitons are
unstable (see Eq. (\ref{crit}) below), while the application of the
management creates a stability region for them in the corresponding
parameter space. Another application of the management to $\mathcal{PT}$%
-symmetric solitons was recently elaborated in terms of the single NLSE,
with a localized complex potential, satisfying condition (\ref{star}) and
subject to cosinusoidal modulation \cite{AF}.

The aim of the present work is to explore the stabilization and dynamical
control of solitons in the $\mathcal{PT}$-symmetric nonlinear coupler by
means of the management applied in a general form, which combines constant
and periodically varying terms in the LG and ICC coefficients. Stability
regions for $\mathcal{PT}$-symmetric solitons are identified by means of
systematic simulations, and also in an analytical form, with the help of the
adiabatic approximation, in the case of the long-period management format.
While the straightforward addition of the management to the system modeling
the $\mathcal{PT}$-symmetric nonlinear coupler leads to shrinkage of the
stability area, defined in terms of the soliton's amplitude (as can be seen
below in Figs. \ref{fig5a}, \ref{fig6a}, and \ref{fig7a}(a)), the periodic
modulation makes it possible to find stable soliton in new situations -- in
particular, in those when the average value of the gain and loss is zero ($%
\gamma _{0}=0$, in terms of Eqs. (\ref{uv})), as shown below in Figs. \ref%
{fig1a}(a), \ref{fig4a}(a), and \ref{fig5a}(d), as well as in the case when
the ICC coefficient may periodically change its sign, which corresponds to $%
\delta >1$, in terms of Eqs. (\ref{uv}) (see Figs. \ref{fig5a}(a-c), \ref%
{fig6a}, \ref{fig7a}(a), and \ref{fig8a} below). Collisions between moving
stable solitons are also considered, by dint of direct simulations.

The rest of the paper is arranged as follows. The model is introduced in
Section 2. The main results, both numerical and analytical, which determine
stability regions for the $\mathcal{PT}$-symmetric solitons, are presented
in Section 3. Collision between stable solitons are addressed in Section 4.
The paper is concluded by Section 5.

\section{The model}

We consider the propagation of optical or matter waves in the dual-core
system described by coupled NLSEs for wave amplitudes $u(z,t)$ and $v(z,t)$
in the cores which carry, severally, gain and loss:

\begin{eqnarray}
i\frac{\partial u}{\partial z}+\frac{1}{2}\frac{\partial ^{2}u}{\partial
t^{2}}+|u|^{2}u-i\left[ \gamma _{0}+\gamma _{1}\sin \left( \frac{2\pi }{L}%
z\right) \right] u+\left[ 1+\delta \sin \left( \frac{2\pi }{L}z+\varphi
\right) \right] v &=&0,  \notag \\
&&  \label{uv} \\
i\frac{\partial v}{\partial z}+\frac{1}{2}\frac{\partial ^{2}v}{\partial
t^{2}}+|v|^{2}v+i\left[ \gamma _{0}+\gamma _{1}\sin \left( \frac{2\pi }{L}%
z\right) \right] v+\left[ 1+\delta \sin \left( \frac{2\pi }{L}z+\varphi
\right) \right] u &=&0.  \notag
\end{eqnarray}%
In terms of optics, $z$ is the propagation distance, while $t$ is the
reduced time in the optical model realized in the temporal domain, as a
dual-core optical fiber \cite{Wabnitz2,Peng}, or the transverse coordinate
in a dual-core planar waveguide, which represents the coupler in the spatial
domain. The group-velocity-dispersion and Kerr coefficients in Eq. (\ref{uv}%
) are scaled to be one, assuming that the dispersion has the anomalous sign,
which is necessary for maintaining bright solitons; in the spatial domain,
the same term represents paraxial diffraction.

The management format, applied to the ICC and LG coefficients, includes
constant terms, with the constant part of the former parameter scaled to be $%
1$, and $\gamma _{0}<1$ being the constant part of the latter one. The ICC
and LG\ modulation amplitudes are, respectively, $\delta $ and $\gamma _{1}$%
, which may be introduced with a phase shift, $\varphi $, and $L$ is the
modulation period. Because swapping the two cores of the coupler represents
the spatial reflection in the present setting, and the propagation distance
is the evolution variable in guided-wave-propagation models, Eqs. (\ref{uv})
are invariant with respect to the $\mathcal{PT}$ transformation, $\left(
u,v\right) \rightarrow \left( v^{\ast },u^{\ast }\right) ,z\rightarrow L/2-z$%
, the shift of $z$ by $L/2$ being a specific feature added by the management
(in other words, $z-L/4$ plays the role of time in the $\mathcal{T}$
transformation).

The dissipative coefficients and $\delta $ are defined to be non-negative,
without the loss of generality:
\begin{equation}
\gamma _{0}\geq 0,~\gamma _{1}\geq 0,~\delta \geq 0  \label{>0}
\end{equation}%
(values $\delta <0$ and $\varphi $ are tantamount to $-\delta $ and $\varphi
+\pi $). Note, however, that the full local LG and ICC coefficients, i.e., $%
\gamma _{0}+\gamma _{1}\sin \left( 2\pi z/L\right) $ and $1+\delta \sin
\left( 2\pi z/L\right) $, respectively, may take negative values -- in
particular, because we will consider, among others, the cases of $\gamma
_{0}=0$ and $\delta >1$.

In the absence of the management, $\gamma _{1}=\delta =0$, a family of exact
soliton solutions to Eq. (\ref{uv}) can be easily found \cite{Driben,Sukho},
provided that $\gamma _{0}\leq 1$:
\begin{equation}
v\left( z,t\right) =\left( i\gamma _{0}\pm \sqrt{1-\gamma _{0}^{2}}\right)
u\left( z,t\right) ,  \label{U}
\end{equation}%
\begin{equation}
v\left( z,t\right) =A\exp \left[ i\left( A^{2}/2\pm \sqrt{1-\gamma _{0}^{2}}%
\right) z\right] \mathrm{sech}\left( At\right) ,  \label{sol}
\end{equation}%
where $A$ is an arbitrary amplitude, and signs $+$ and $-$ correspond to the
$\mathcal{PT}$-symmetric and antisymmetric solitons, respectively, which are
so named \cite{Driben} because they correspond, respectively, to the usual
symmetric and antisymmetric solitons in the usual coupler's model in the
absence of the gain and loss ($\gamma _{0}=0$; note that more general
compound solitons, with an arbitrary phase difference between the two
components, are not possible, because the $\mathcal{P}$ transformation has
only two eigenvalues, $+1$ and $-1$). Note that solution (\ref{U}) implies%
\begin{equation}
\left\vert u\left( z,t\right) \right\vert ^{2}=\left\vert v\left( z,t\right)
\right\vert ^{2},  \label{=}
\end{equation}%
which maintains equilibrium between the gain and loss.

The stability region for the exact symmetric solitons in the static model ($%
\gamma _{1}=\delta =0$), given by Eqs. (\ref{U}) and (\ref{sol}),\ can also
be found in an exact form \cite{Driben,Sukho}: they are stable if the
squared amplitude takes values
\begin{equation}
A^{2}\leq A_{\mathrm{crit}}^{2}\left( \gamma _{0}\right) =(4/3)\sqrt{%
1-\gamma _{0}^{2}},  \label{crit}
\end{equation}%
while the antisymmetric solitons are completely unstable (although their
instability may be very weak, depending on the parameters). For this reason,
antisymmetric solitons are not considered in detail below below (they may be
made stable in a discrete version of Eqs. (\ref{uv}) \cite{Barash-discr}).

As mentioned above, an essential difference of the $\mathcal{PT}$-symmetric
system (\ref{uv}) from its conservative counterpart, with $\gamma _{0}=0$,
is that, at $A^{2}>A_{\mathrm{crit}}^{2}$, unstable $\mathcal{PT}$-symmetric
solitons are not replaced by stable asymmetric ones (cf. works \cite{Wabnitz}%
-\cite{Peng}, where asymmetric solitons are considered in the conservative
system), because asymmetric states, that do not obey condition (\ref{=}),
cannot maintain the LG balance (i.e., asymmetric solitons cannot exist in
the case of $\gamma _{0}>0$). As a result, unstable\ $\mathcal{PT}$%
-symmetric soliton suffer blowup, similar to what is shown below in Fig. \ref%
{fig2a}(c).

\section{Stabilization and dynamical control of solitons by the management}

\subsection{Numerical results}

We focus on the case of zero phase shift between the variations of the LG
and ICC coefficients, i.e., $\varphi =0$ in Eq. (\ref{uv}), as the
management format with $\varphi \neq 0$ (in particular, $\varphi =\pi $,
which, as mentioned above, is tantamount to taking $\delta <0$) leads to
strong instability. The stability of the solitons under the action of the
management was identified from sufficiently long direct simulations, with
the input taken in the form of Eqs. (\ref{U}) and (\ref{sol}) at $z=0$ and a
given value of $\gamma _{0},$ while $A$ was varied, to collect systematic
results for the stability of the solitons with different amplitudes, cf. Eq.
(\ref{crit}). The simulations were carried out by means of the split-step
numerical algorithm, similar to those employed in Refs. \cite{Driben} and
\cite{Sukho}. A rigorous study of the stability against small perturbations,
which makes it necessary to solve linearized equations around the
periodically varying solution, is a challenging task, which we do not tackle
here.

\begin{figure}[tbp]
\centering{\includegraphics[scale=0.65]{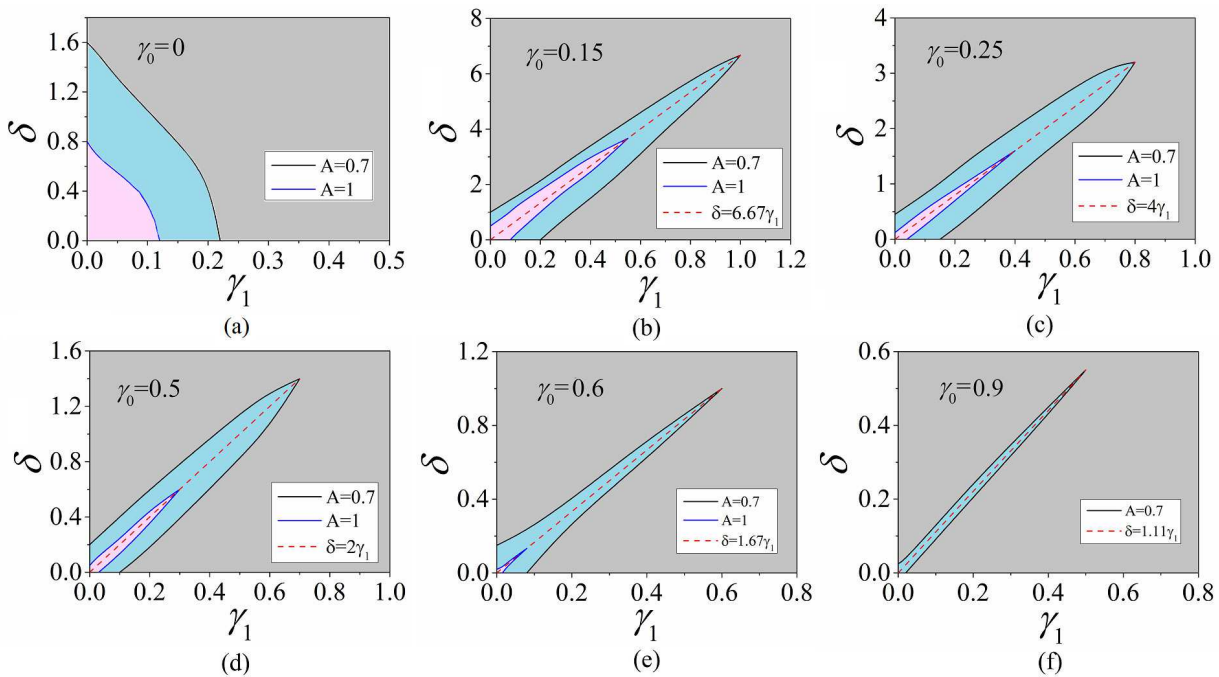}}
\caption{(Color online) Stability charts for $\mathcal{PT}$-symmetric
solitons under the action of the management with a fixed value of the
period, $L=\protect\pi /3$. In panels (a) to (f), the constant part of the
LG coefficient is $\protect\gamma _{0}=0,0.15,0.25,0.5,0.6,0.9$,
respectively. The solitons, with initial amplitudes $A=0.7$ and $1$ in Eq. (%
\protect\ref{sol}), are stable, respectively, in the blue and pink areas in
the plane of the modulation amplitudes, $(\protect\gamma _{1},\protect\delta %
)$ (in the pink areas, both $A=0.7$ and $1$ produce stable solitons). The
straight dashed line denotes the optimum-stability relation (\protect\ref{//}%
), see the text. The simulations produce unstable evolution in the ambient
gray region. In panel (f), there is no stability area for $A=1$, in
agreement with Eq. (\protect\ref{crit}), which yields, in this case, $A_{%
\mathrm{crit}}^{2}\left( 0.9\right) \approx 0.58$. }
\label{fig1a}
\end{figure}

A combination of panels displayed in Fig. \ref{fig1a} show stability areas
for the solitons with different values of the constant part of the LG
coefficient, $\gamma _{0}$, for a fixed management period, $L=\pi /3$, and
two different values of amplitude $A$ in the input expression (\ref{sol}).
Naturally, the stability areas are larger for smaller $A$ (similar to what
is predicted by Eq. (\ref{crit}) in the absence of the management) , and
they shrink with the increase of the modulation amplitudes, $\gamma _{1}$
and $\delta $. It is clearly seen that the strongest stability is provided
by the management scheme in which the ratio of $\delta $ and $\gamma _{1}$
is the same as the ratio of their counterparts in the constant parts of the
ICC and LG coefficients (in other words, the management is applied\textit{\
coherently }with the static part of the system):

\begin{equation}
\delta /\gamma _{1}=1/\gamma _{0}.  \label{//}
\end{equation}%
This finding is explained by the fact that, when relation (\ref{//}) holds,
Eqs. (\ref{uv}) admit \emph{exact solutions} for $\mathcal{PT}$-symmetric
and antisymmetric solitons:
\begin{equation}
v\left( z,t\right) =A\exp \left[ i\left( A^{2}/2\pm \sqrt{1-\gamma _{0}^{2}}%
\right) z\mp i\frac{L\gamma _{1}\sqrt{1-\gamma _{0}^{2}}}{2\pi \gamma _{0}}%
\cos \left( \frac{2\pi }{L}z\right) \right] \mathrm{sech}\left( At\right) ,
\label{phase}
\end{equation}%
cf. Eq. (\ref{sol}), with $u\left( z,t\right) $ given by exactly the same
equation (\ref{U}) as above. Of course, stability conditions for these exact
solutions cannot be found in the same simple form (\ref{crit}) which is
valid in the absence of the management.

A principally different case is one shown in Fig. \ref{fig1a}(a), which
corresponds to $\gamma _{0}=0$ (i.e., the static system is the conservative
one). In this case, relation (\ref{//}) does not exist, and, accordingly,
the shape of the stability area is completely different from those displayed
in panels (b)-(f).

\begin{figure}[tbp]
\centering{\includegraphics[scale=0.3]{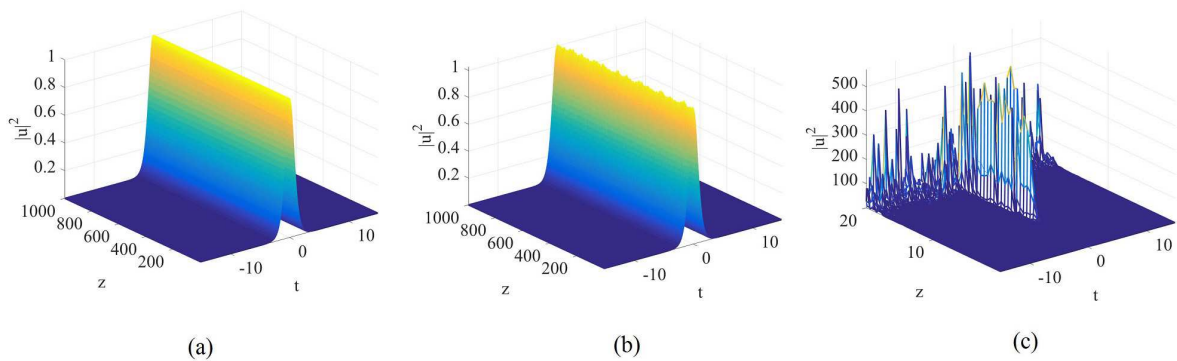}}
\caption{(Color online) Typical examples of the evolution of stable and
unstable solitons under the action of the management, for fixed parameters $%
(L,\protect\gamma _{0},\protect\gamma _{1},A)=(\protect\pi /3,0.5,0.3,1)$,
while $\protect\delta $ is varied. In panel (a), with $\protect\delta =0.6$,
which satisfies the optimum-stability relation (\protect\ref{//}), the
soliton's shape remains unchanged in the course of the propagation, in
agreement with Eq. (\protect\ref{phase}). In panel (b), $\protect\delta %
=0.64 $, which does not meet condition (\protect\ref{//}) (this $\protect%
\delta $ corresponds to $\protect\gamma _{0}\left( \protect\delta /\protect%
\gamma _{1}\right) \approx 1.07$, instead of $1$), but belongs to an edge of
the stability area in Fig. \protect\ref{fig1a}(d). In this case, shape
oscillations are small but visible. Panel (c) is an example of the evolution
of an unstable soliton with $\protect\delta =0.1$, which lies deep in the
gray area in Fig. \protect\ref{fig1a}(d). This soliton blows up due to the
imbalance between the gain and loss (note the difference in the vertical
scales between panels (a,b) and (c)).}
\label{fig2a}
\end{figure}

\begin{figure}[tbp]
\centering{\includegraphics[scale=0.15]{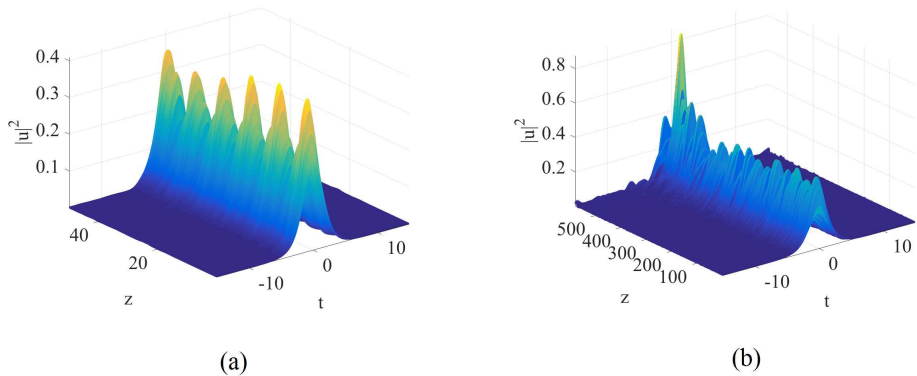}}
\caption{(Color online) A typical example of the evolution of a moderately
unstable breather, for parameters $(L,\protect\gamma _{0},\protect\gamma %
_{1},\protect\delta ,A)=(\protect\pi ,0.9,0.2,0.12,0.5)$. The breather
remains quasi-stable over a short propagation distance in panel (a), but
eventuallybecomes unstable in the course of the further evolution, as shown
in panel (b). }
\label{fig3a}
\end{figure}

The evolution of a stable soliton which satisfies constraint (\ref{//}) is
displayed in Fig. \ref{fig2a}(a), showing that it keeps a constant shape, in
exact agreement with Eqs. (\ref{phase}) and (\ref{U}). On the other hand, in
the case when the management parameters deviate from condition (\ref{//}), a
typical example of the evolution of stable solitons shows small but visible
fluctuations in Fig. \ref{fig2a}(b). On the contrary, unstable solitons blow
up due to the failure of the LG balance, see an example of the quick onset
of the blowup in Fig. \ref{fig2a}(c), for parameters chosen deep in the
instability area. Moderately unstable solitons develop into breathers, which
evolve as quasi-stable states (cf. similar dynamical modes reported in Ref.
\cite{Sukho}), but eventually they are destroyed by the collapse, as shown
in Fig. \ref{fig3a}.

\begin{figure}[tbp]
\centering{\includegraphics[scale=0.65]{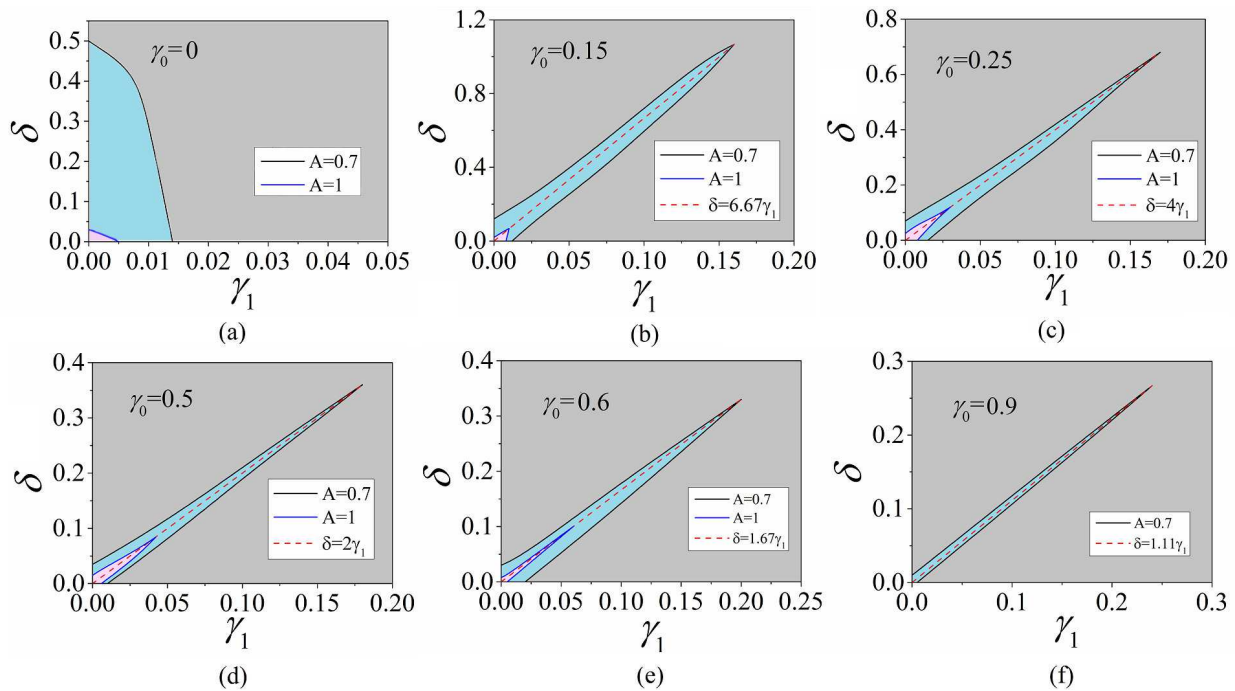}}
\caption{(Color online) The same as in Fig. \protect\ref{fig1a}, but for the
management period which is three times as large, $L=\protect\pi $. The
structure of the stability domains remains similar, but their overall size
essentially decreases, in comparison with the case of $L=\protect\pi /3$. In
particular, there is no stability area for $A=1$ in panel (f), for the same
reason as in Fig. \protect\ref{fig1a}(f).}
\label{fig4a}
\end{figure}

Similar results, pertaining to a larger management period, $L=\pi $, are
presented in Fig. \ref{fig4a}. It is seen that they are qualitatively
similar to those in Fig. \ref{fig1a}, but the large period supports smaller
stability areas, both in the case of $\gamma _{0}=0$ and $\gamma _{0}>0$.
Note that the same condition (\ref{//}) determines the condition of the
optimum stability in this case too, when exact $\mathcal{PT}$-symmetric
solitons are given by Eqs. (\ref{U}) and (\ref{phase}).

\begin{figure}[tbp]
\centering{\includegraphics[scale=0.6]{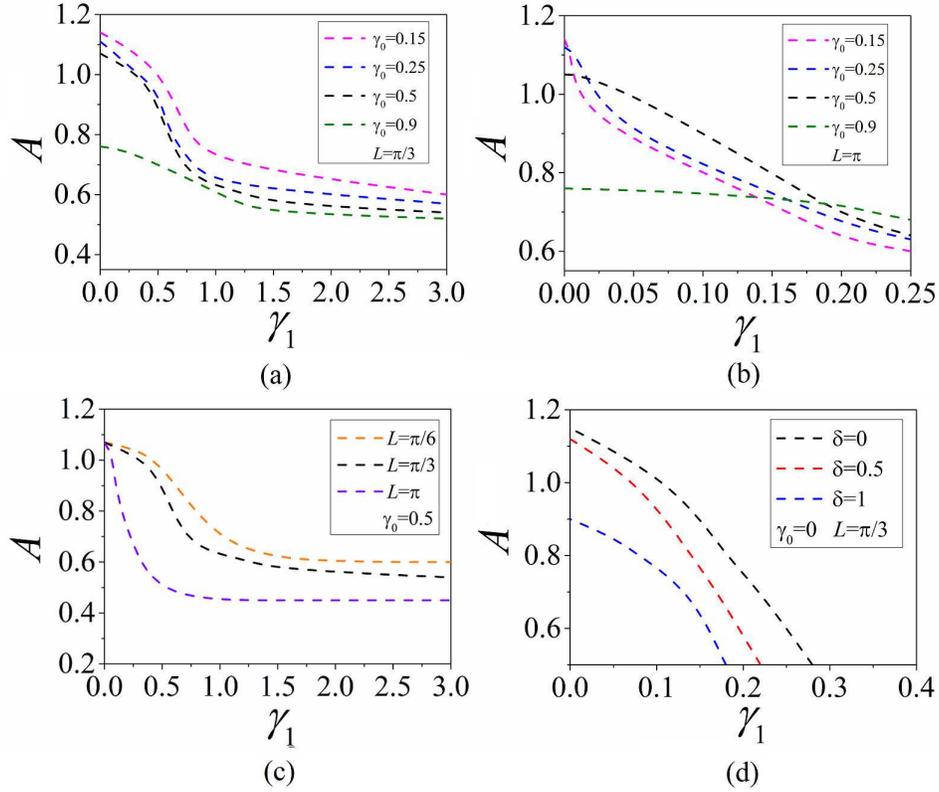}}
\caption{(Color online) Panels (a,b,c): imposing the optimum-stability
condition (\protect\ref{//}), i.e., $\protect\delta =\protect\gamma _{1}/%
\protect\gamma _{0}$, stability boundaries for $\mathcal{PT}$-symmetric
solitons are displayed in the plane of $(\protect\gamma _{1},A)$ (recall $A$
is the input's amplitude, according to Eq. (\protect\ref{sol})). The
solitons are stable beneath boundaries shown in the panels. Widely different
ranges of parameter $\protect\gamma_1$, displayed in panels (a), (c) and
(b), (d), correspond to the fact that characteristic values of this
parameter are indeed strongly different in the respective cases. In (a) and
(b), the management period is fixed, respectively, at $L=\protect\pi /3$ and
$L=\protect\pi $, and the results are presented for a set of different
values of $\protect\gamma _{0}$, cf. Figs. \protect\ref{fig1a} and \protect
\ref{fig4a}. In panel (c), $\protect\gamma _{0}=0.5$ is fixed, and a set of
the stability boundaries are displayed for different periods $L$. Panel (d)
is plotted for $\protect\gamma _{0}=0$, for which condition (\protect\ref{//}%
) is irrelevant. In this case, the stability boundaries are presented for
fixed $L=\protect\pi /3$ and three different values of the amplitude of the
ICC modulation: $\protect\delta =0,0.5,$ and $1$.}
\label{fig5a}
\end{figure}

Another essential summary of the results is presented in Fig. \ref{fig5a},
in the form of stability maps in the plane of $\left( \gamma _{1},A\right) $%
, while $\delta $ is linked to $\gamma _{1}$ by the optimum-stability
condition (\ref{//}). At $\gamma _{1}=0$ (hence $\delta =0$ too), i.e., in
the absence the management, the largest values of $A$ admitting the
stability in Figs. \ref{fig5a}(a) and (b) are precisely the same as
predicted analytically by Eq. (\ref{crit}) for the static $\mathcal{PT}$%
-symmetric coupler.

A natural trend evidenced by Figs. \ref{fig5a}(a) and (b) is that the
stability limit, given by the largest value of $A$ up to which the solitons
persist, decreases with the increase of the modulation strength, $\gamma
_{1} $. A noteworthy feature observed by Fig. \ref{fig5a}(b) is that, for a
relatively large management period, $L=\pi $, the stability boundary for
larger $\gamma _{0}$ may be located higher, in terms of $A$, than its
counterpart for smaller $\gamma _{0}$, see, for instance, the boundaries for
$\gamma _{0}=0.15$ and $\gamma _{0}=0.5$. This feature is counter-intuitive,
as the exact result (\ref{crit}) for the solitons in the static model
demonstrates monotonic decay of $A_{\mathrm{crit}}$ with the increase of $%
\gamma _{0}$. An explanation for this point is that, at smaller $\gamma _{0}$%
, the stability is more sensitive to changes of period $L$. The analysis of
the situation for the long-period modulations with large $L$ is presented in
the next section, with the help of the adiabatic approximation.

Lastly, Fig. \ref{fig5a}(d) demonstrates that, in the case of $\gamma _{0}=0$%
, when the constant term is absent in the LG coefficient (hence Eq. (\ref{//}%
) is irrelevant), the increase of the modulation amplitudes of both the LG
and ICC terms, i.e., $\gamma _{1}$ and $\delta $, naturally causes shrinkage
of the stability area.

\subsection{The adiabatic approximation}

The long-period management may be considered as adiabatic under the
condition that the period is much larger than the intrinsic period of phase
oscillations of soliton (\ref{sol}):
\begin{equation}
L\gg 4\pi /A^{2}.  \label{L}
\end{equation}%
This condition makes it possible to separate the scales of the slow
evolution of stable solitons, driven by the management, and their rapid
phase oscillations, which correspond to the nearly-constant values of the
parameters

The adiabatic limit allows one to approximately transform Eq. (\ref{uv})
into equations with constant coefficients, by defining%
\begin{eqnarray}
t &\equiv &\frac{\tilde{t}}{\sqrt{1+\delta \sin \left( \frac{2\pi }{L}%
z\right) }},  \label{tau} \\
z &\equiv &\frac{\tilde{z}}{1+\delta \sin \left( \frac{2\pi }{L}z\right) },
\label{z} \\
\left( u,v\right)  &\equiv &\left( \tilde{u},\tilde{v}\right) \sqrt{1+\delta
\sin \left( \frac{2\pi }{L}z\right) },  \label{prime}
\end{eqnarray}%
where it is assumed that $\gamma _{1}$ and $\delta $ are related by Eq. (\ref%
{//}), to address the case which is most relevant for the stability
analysis. The substitution of variables (\ref{tau})-(\ref{prime}) leads, in
the first approximation, to the following equations replacing Eq. (\ref{uv}%
):
\begin{eqnarray}
i\frac{\partial \tilde{u}}{\partial \tilde{z}}+\frac{1}{2}\frac{\partial ^{2}%
\tilde{u}}{\partial \tilde{t}^{2}}+|\tilde{u}|^{2}\tilde{u}-i\gamma _{0}%
\tilde{u}+\tilde{v} &=&0,  \notag \\
&&  \label{tilde} \\
i\frac{\partial \tilde{v}}{\partial \tilde{z}}+\frac{1}{2}\frac{\partial ^{2}%
\tilde{v}}{\partial \tilde{t}^{2}}+|\tilde{v}|^{2}\tilde{v}+i\gamma _{0}%
\tilde{v}+\tilde{u} &=&0.  \notag
\end{eqnarray}%
This approximate transformation is relevant under condition
\begin{equation}
\delta <1,  \label{<1}
\end{equation}%
which is necessary to secure condition $1+\delta \sin \left( 2\pi z/L\right)
>0$; otherwise, the transformation given \ by (\ref{tau})-(\ref{prime})
becomes singular. Taking into regard the currently imposed relation (\ref{//}%
), Eq. (\ref{<1}) may also be written as $\gamma _{1}<\gamma _{0}$.

Being tantamount in their form to Eq. (\ref{uv}), equations (\ref{tilde})
produce solutions in the form of (\ref{U}) and (\ref{sol}), which, in turn,
are stable under the accordingly transformed criterion (\ref{crit}). The
critical condition corresponds to the largest amplitude, in terms of the
transformed fields (\ref{prime}), which is $\tilde{A}_{\mathrm{crit}%
}^{2}=A^{2}/\left( 1-\delta \right) $, attained at $\sin \left( 2\pi
z/L\right) =-1$. In terms of the original notation, the respective
approximate stability criterion for the slowly varying solitons takes the
form of
\begin{equation}
A_{0}^{2}<\left( A_{0}^{2}\right) _{\mathrm{crit}}=\frac{4}{3}\sqrt{1-\gamma
_{0}^{2}}\left( 1-\delta \right) \equiv \frac{4}{3}\sqrt{1-\gamma _{0}^{2}}%
\left( 1-\frac{\gamma _{1}}{\gamma _{0}}\right)  \label{<}
\end{equation}%
(Eq. (\ref{//}) is used to write the result in Eq. (\ref{<}) in two
equivalent forms).

\begin{figure}[tbp]
\centering{\includegraphics[scale=0.65]{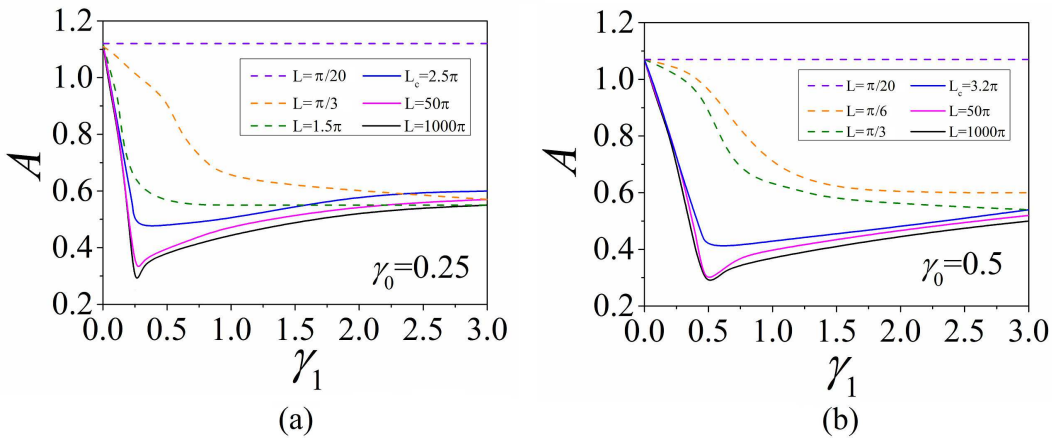}}
\caption{(Color online) Stability boundaries in the plane of $\left( \protect%
\gamma _{1},A\right) $ under condition (\protect\ref{//}), for six very
different values of the management period, from $L=\protect\pi /20$ to $%
L=1000\protect\pi $, and two different values of $\protect\gamma _{0}$. The
stability area shrinks but does not vanish with the increase of $L$. The
shape of the stability boundary becomes non-monotonous (the minimum point
appears on it) at $L\geq L_{\mathrm{c}}$. The solid and dashed lines
designate the non-monotonous and monotonous stability boundaries,
respectively.}
\label{fig6a}
\end{figure}

The global picture of the transformation of the stability boundary in the
plane of $\left( \gamma _{1},A\right) $ is illustrated in Fig. (\ref{fig6a}%
), by showing it for six different management periods, which cover the range
of four order of magnitude (from $L=\pi /20$ to $L=1000\pi $), and two
values $\gamma _{0}=0.25,0.5$. It is observed that, in each case, there is a
critical value, $L_{\mathrm{c}}$, such that dependence $A\left( \gamma
_{1}\right) $ along the stability boundary is monotonous at $L<L_{\mathrm{c}%
} $, and non-monotonous at $L>L_{\mathrm{c}}$.

\begin{figure}[tbp]
\centering{\includegraphics[scale=0.65]{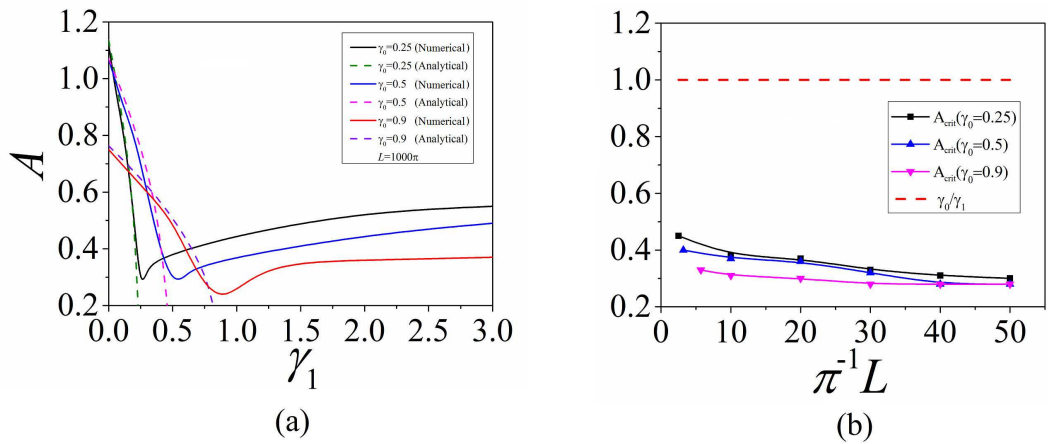}}
\caption{(Color online) (a) Stability boundaries in the plane of $\left(
\protect\gamma _{1},A\right) $ under condition (\protect\ref{//}), at a very
large fixed management period, $L=1000\protect\pi $, and three different
values of $\protect\gamma _{0}$. Solid and dashed lines represent,
severally, numerical results and analytical ones predicted by Eq. (\protect
\ref{<}). (b) Value $A_{\mathrm{crit}}$ of the amplitude at the minimum
point vs. $L$, at three different values of $\protect\gamma _{0}$. The
horizontal dashed line shows the value of $\protect\gamma _{0}/\protect%
\gamma _{1}$ at the minimum point, confirming that it exactly corresponds to
$\protect\gamma _{1}=\protect\gamma _{0}$, as predicted by the analytical
approximation.}
\label{fig7a}
\end{figure}

Further, the analytical prediction given by Eq. (\ref{<}) is compared to the
numerically found stability boundaries, for a very large period, $L\equiv
1000\pi $, in Fig. \ref{fig7a}(a). It is seen that the prediction is close
to the numerical counterparts in the region of $\gamma _{1}<\gamma _{0}$. At
$\gamma _{1}=\gamma _{0}$ the analytically predicted critical amplitude
vanishes in Eq. (\ref{<}), and it ceases to exist at $\gamma _{1}>\gamma _{0}
$, i.e., in the case when the sign of the total LG periodically changes,
according to Eq. (\ref{uv}). In fact, the analytical approximation breaks
down close to $\gamma _{1}=\gamma _{0}$ (i.e., close to $\delta =1$, see Eq.
(\ref{<1})), as mentioned above. In fact, the numerically found critical
amplitude does not vanish at point $\gamma _{1}=\gamma _{0}$, but, instead,
it attains a finite minimum value. At $\gamma _{1}>\gamma _{0}$, the
stability area still exists, slowly expanding with the increase of $\gamma
_{1}/\gamma _{0}\equiv \delta $. This trend is a natural one, as the
increase of the absolute value of the ICC suppresses the symmetry-breaking
instability driven by the self-focusing nonlinearity \cite{Wabnitz}-\cite{Peng}.

At all values of $\gamma _{0}$, the minimum of $A_{\mathrm{crit}}$ is
attained exactly at $\gamma _{1}=\gamma _{0}$, in full agreement with the
analytical prediction. This fact is confirmed by the numerically generated
value of $\gamma _{0}/\gamma _{1}$ at the minimum point, which is shown,
versus $L$, by the dashed line in Fig. \ref{fig7a}(b). As concerns value $%
A_{\min }$ of the critical amplitude at the minimum point, it slowly
decreasing with the increase of $L$, as is also shown in Fig. \ref{fig7a}(b).

\section{Collisions of solitons}

Because Eq. (\ref{uv}) keeps the Galilean invariance, a stable soliton can
be set in motion by applying kick $\eta $ to it, i.e., multiplying the
quiescent solution $\left\{ u_{0}\left( z,t\right) ,v_{0}\left( z,t\right)
\right\} $ by $\exp \left( i\eta t\right) $. As a result, it will be
transformed into a moving one,
\begin{equation}
\left\{ u_{\eta },v_{\eta }\right\} =\left\{ u_{0}\left( z,t-cz\right)
,v_{0}\left( z,t-cz\right) \right\} \exp \left( -\left( i/2\right)
c^{2}z\right) ,~c=\eta .  \label{kick}
\end{equation}%
This fact suggests to simulate collisions between initially separated
solitons, boosted in opposite directions by kicks $\pm \eta $, cf. Refs.
\cite{Driben} and \cite{Sukho}.

\begin{figure}[tbp]
\centering{\includegraphics[scale=0.4]{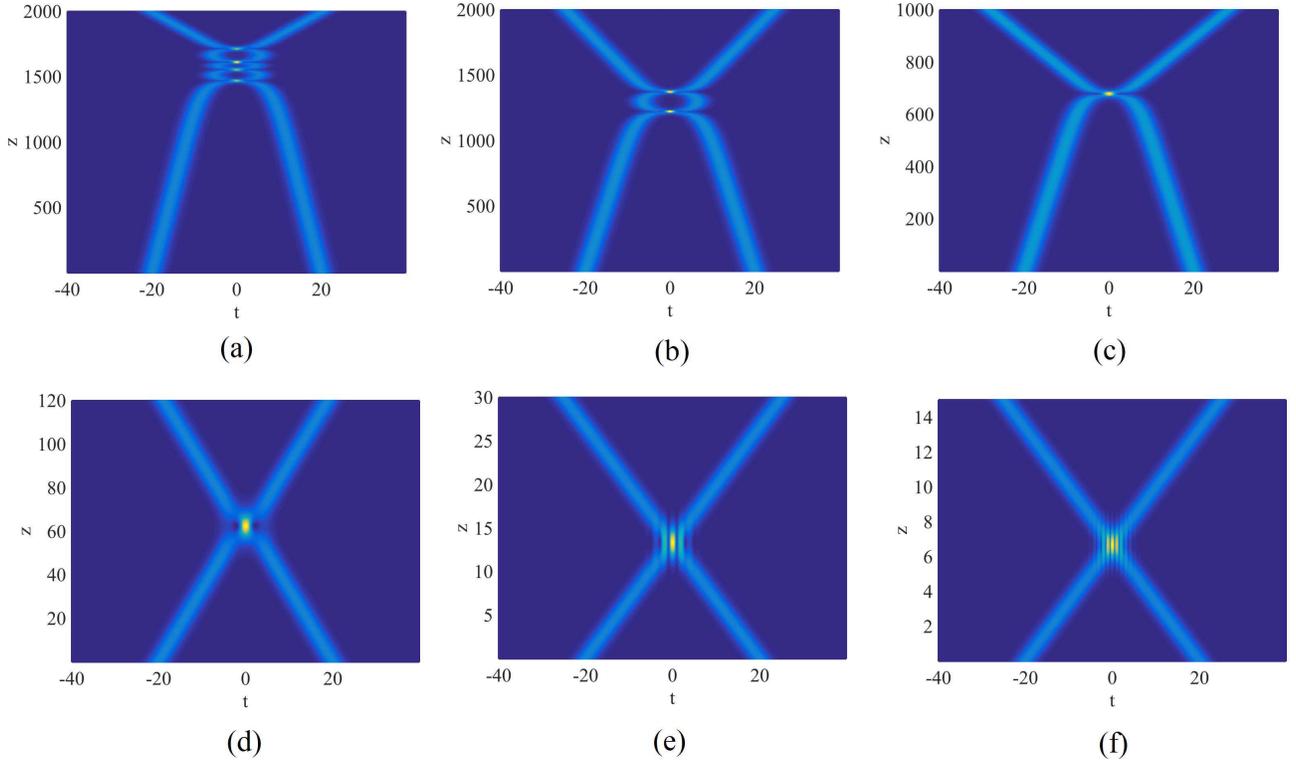}}
\caption{(Color online) Typical examples of collisions between identical
stable $\mathcal{PT}$-symmetric solitons moving in opposite directions under
the action of kicks $\pm \protect\eta $, see Eq. (\protect\ref{kick}). The
solitons' parameters are $(L,\protect\gamma _{0},\protect\gamma _{1},\protect%
\delta ,A)=(\protect\pi /3,0.5,1,2.05,0.5)$. From panel (a) to (f), the
kicks applied to the soliton pairs are $\protect\eta =\pm 0.008,\pm 0.01,\pm
0.02,\pm 0.3,\pm 1.5,\pm 3$, respectively. }
\label{fig8a}
\end{figure}

A typical set of collisions, simulated for a set of different values of the
kicks, is displayed in Fig.(\ref{fig8a}). Panels(a) to (f) show different
outcomes of the collisions, produced by increasing values of $\eta $. In all
the cases, the colliding solitons eventually separate. However, in panels
(a) and (b) slowly moving soliton pairs form quasi-bound states, in which
they perform several oscillations before re-emerging with larger values of
opposite velocities $\pm c$ (see Eq. (\ref{kick})) than they had prior to
the collision. The formation of the intermediate bound state resembles the
effect which was previously found in simulations of soliton-soliton
collisions in other nonintegrable models \cite{Campbell1,Campbell2,RMP}.
Fast moving solitons, boosted by stronger kicks, pass through each other
elastically (panels (d,e,f)), which is typical for collisions between
solitons in conservative systems \cite{RMP}, and remains true in the present
$\mathcal{PT}$-symmetric one.

\begin{figure}[tbp]
\centering{\includegraphics[scale=0.4]{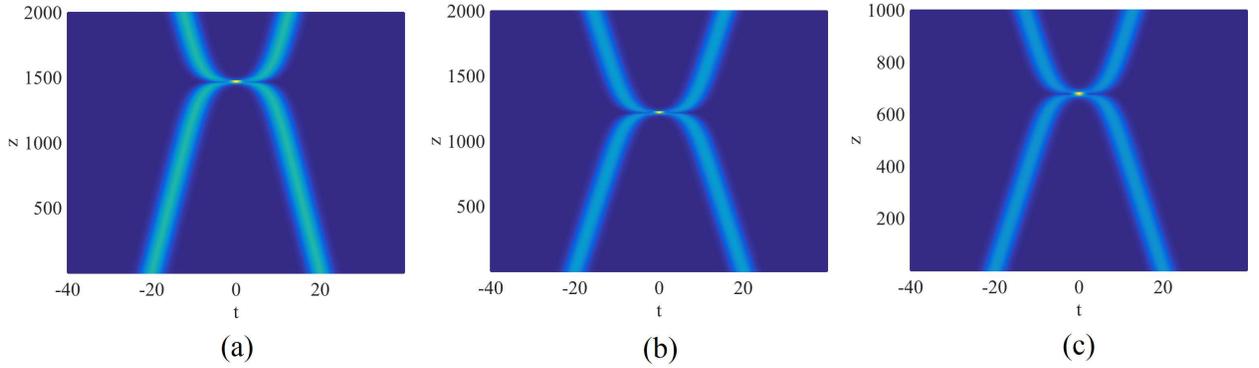}}
\caption{(Color online) The same as in the top row of Fig. \protect\ref%
{fig8a}, but in the absence of the management, i.e., with $\protect\gamma %
_{1}=\protect\delta =0$, while other parameters, including the kicks, keep
the same values. }
\label{fig9a}
\end{figure}

Lastly, Fig. \ref{fig9a} displays simulations of the collisions with the
same values of $L,\gamma _{0}$, and $A$, under the action of the same kicks
as in the top row in Fig. \ref{fig8a}, but in the absence of the management,
i.e., for $\gamma _{1}=\delta =0$. It is clearly seen that the collisions
are completely elastic (similar to those simulated in \cite{Driben}), and
the intermediate quasi-bound states do not emerge. Thus, the presence of the
management accounts for the creation of the those states. For larger kicks,
corresponding to the bottom row in Fig. \ref{fig8a}, the collisions remain
the same (elastic) as displayed in Fig. \ref{fig8a}.

As said above, in this work we do not address $\mathcal{PT}$-antisymmetric
solitons, which correspond to the bottom signs in Eqs. (\ref{U}), (\ref{sol}%
) and (\ref{phase}), as they are unstable even in the absence of the
management. The instability of some antisymmetric solitons being weak, it is
possible to consider their collisions too, with each other or with $\mathcal{%
PT}$-symmetric counterparts. As shown in Ref. \cite{Sukho}, in the latter
case the collision may excite intrinsic oscillations in the solitons. The
consideration of this case is beyond the scope of the present work.

\section{Conclusion}

The objective of this work is to generalize the known model of the $\mathcal{%
PT}$-symmetric coupler, based on linearly coupled waveguides with the
intrinsic cubic nonlinearity and equal gain and loss coefficients carried by
the guiding cores. The generalization introduces \textquotedblleft
management", which makes the LG (loss-gain) and ICC (inter-core-coupling)
coefficients periodically varying functions of the evolutional variable. The
model may be realized in optics, in the temporal and spatial domains alike.
Stability of $\mathcal{PT}$-symmetric solitons and possibilities of applying
the dynamical control to them by means of the management are explored by
means of systematic simulations, and also analytically in the adiabatic
approximation, which corresponds to the long-period limit. The stability is
strongest when the ratio of the amplitudes of the modulation of the LG and
ICC coefficients is equal to its counterpart for the constant parts of the
same coefficients. In the latter case, an exact solution is found for $%
\mathcal{PT}$-symmetric solitons. Collisions between moving solitons were
briefly considered too.

A challenging possibility for the development of the present analysis is to
develop analysis of the management for solitons in two-dimensional $\mathcal{%
PT}$-symmetric systems.

\begin{acknowledgments}
This work was supported, in part, by the Israel Science Foundation through
grant No. 1287/17. We appreciate a helpful discussion with Zhaopin Chen (Tel
Aviv University) and technical assistance provided by Ms. Shiyue Liu
(Chinese University of Hong Kong).
\end{acknowledgments}

\end{document}